\documentclass[letterpaper, final, twocolumn, prl, showpacs]{revtex4}

\usepackage{mathptmx}

\usepackage{amssymb}

\usepackage{graphicx}

\newcommand{\mean}[1]{\langle #1 \rangle}

\newcommand{\defeq}{:=}

\newcommand{\x}{\mathbf{x}}
\renewcommand{\v}{\mathbf{v}}
\newcommand{\msd}{\mean{r(t)^2}}

\renewcommand{\P}{\mathbb{P}}

\graphicspath{{figs/}}

\begin{document}


\title{Normal diffusion in 
crystal structures and higher-dimensional
billiard models with gaps}

\author{David P.~Sanders}
\email{dps@fciencias.unam.mx}
\affiliation{Departamento de F\'isica, Facultad de Ciencias, Universidad
Nacional Aut\'onoma de M\'exico, 
M\'exico D.F.,
04510 Mexico}

\date{\today}



\begin{abstract}
%


We show, both heuristically and numerically, that three-dimensional
periodic Lorentz 
gases---clouds of 
particles scattering off 
crystalline arrays of 
 hard spheres---often exhibit normal
diffusion, 
even when there are \emph{gaps} through which particles can
travel
without ever colliding,
i.e., when the system has an infinite horizon.
This is the case provided that these
gaps 
are not ``too big'', as measured by their dimension.
The results are illustrated 
with 
simulations
 of a simple three-dimensional model 
having different types of diffusive regime,
and are then extended to
higher-dimensional
billiard models, which include hard-sphere fluids.


\end{abstract}

\maketitle


The Lorentz gas is a classical model of transport processes,
in which a cloud
of non-interacting point particles (modelling electrons) undergo
free motion between
elastic collisions with fixed hard spheres (atoms)
\cite{LorentzOriginal1905}.  
It has been much studied as a model
system for which 
the programme of statistical
physics can be carried out in detail: to relate
the known microscopic dynamics to the 
macroscopic behavior of the system, which in this case is diffusive
\cite{GaspardBook1998, SzaszHardBallSystemsAndLorentzGasBook2000, KlagesMicroscopicChaosBook}.


When the scatterers are arranged in a periodic crystal structure, 
the dynamics of this \emph{billiard model} can be reduced to a
single unit cell \cite{GaspardBook1998}.
The curved shape of
the scatterers 
implies that nearby trajectories separate
exponentially fast, so that the
system is hyperbolic (chaotic) and ergodic
\cite{ChernovMarkarianChaoticBilliardsAMS2006}.

In two dimensions, it has been shown 
that the cloud of particles in the periodic Lorentz gas undergoes
normal diffusion, provided that the geometrical \emph{finite horizon}
condition is satisfied: particles cannot travel arbitrarily far without
colliding with a scatterer
\cite{BunimovichSinaiStatPropsPeriodicLorentzCMP1981,
ChernovMarkarianChaoticBilliardsAMS2006}.
%
%
By \emph{normal diffusion}, we mean that the distribution of
particle positions behaves 
like solutions of the diffusion equation; in particular, the
mean-squared displacement (variance) grows asymptotically
linearly in time:
$\msd \sim 2dDt$ 
when
$t \to \infty$,
where 
$r(t)$ is 
the displacement of a particle at time $t$
from its initial position, 
$d$ is the
spatial dimension,  $\mean{\cdot}$ denotes a mean over initial conditions, and
the diffusion coefficient $D$ gives the asymptotic growth rate.

When the horizon is infinite, however, particles can undergo arbitrarily long
free flights along \emph{corridors} in the
structure.
It was long argued 
\cite{FriedmanMartinDecayVAFLorentzPhysLettA84,
ZacherlGeiselAnomDiffnExtendedSinaiBilliardPhysLettA1986,
BleherStatProps2DInfHorizJSP1992}
and has
recently been proved 
\cite{SzaszVarjuLimitLawsRecurrencePlanarLorentzInfiniteHorizonJSP2007}, that
there is then weak superdiffusive
behavior, with $\msd \sim t \log t$, so that the diffusion coefficient no
longer exists.

For \emph{higher-dimensional} periodic Lorentz gases, 
rigorous
results on ergodic properties
\cite{ChernovSinaiErgPropsThreeDimensionalBallsRussMathSurv1987} and diffusive
properties \cite{ChernovStatPropsMultidimPeriodicLorentzGasJSP1994} have been
obtained; recent progress in their analysis has been made
\cite{BalintMultdimSemidispersingSingularitiesAnnHenriPoinc2002,
BalintTothExpDecayMultiDimensional}, including in the limit of small scatterers 
\cite{MarklofDistributionFreePathLengthsPreprint2007}. In particular,
higher-dimensional Lorentz gases are believed to exhibit normal diffusion when
the horizon is finite \cite{ChernovStatPropsMultidimPeriodicLorentzGasJSP1994}.
%

Nonetheless, the study of billiard models in higher
dimensions, especially three dimensions, has
received surprisingly little attention from the physics community, despite
their interest as simple models of transport in
three-dimensional crystals. 
%
This can be attributed to increased simulation times and the difficulty of
visualisation in higher dimensions, but also to an apparent general belief that
the diffusive properties of
higher-dimensional
systems should be analogous to those in the 2D case. Hypercubic Lorentz
gases (with infinite horizon) in $d \le 7$ dimensions were
studied in 
\cite{BouchaudLeDoussalHigherDimensionalLorentzGasJSP1985}, but no strong
conclusions about diffusive properties could be drawn.

 In particular, it was believed that a finite horizon was necessary for a system
to
show normal diffusion, with weak superdiffusion occurring for an infinite
horizon \cite{DettmannLorentzGasReviewInSzasz2000,
ChernovStatPropsMultidimPeriodicLorentzGasJSP1994}. While periodic 
Lorentz gases with finite horizon and disjoint obstacles have been proved to \emph{exist} in any dimension
\cite{HenkZongSegmentsBallPackingsMathematika2000}, 
constructing such a model 
turns out to be a difficult task---we are not aware of 
\emph{any} known explicit examples, even in three dimensions.
Furthermore, crystals of spheres arranged in any Bravais lattice (and in many
other crystal structures) always have small gaps 
which prevent a finite horizon \cite{HeppesLatticePackings1961,
HenkZongSegmentsBallPackingsMathematika2000,
ZongDeepHolesFreePlanesReviewBullAMS2002}.

%



In this Letter, we show, using heuristic arguments and careful numerical
simulations, that in fact
periodic Lorentz gases in three and higher dimensions with infinite
horizon---that is, with \emph{gaps}, or holes, in the structure---can exhibit
\emph{normal} diffusion. The
key
observation is that 
the gaps 
in configuration space,
which are higher-dimensional analogs of the corridors in 2D, can be of
different dimensions. Structures with gaps of the highest
possible dimension exhibit weak superdiffusion, as in the 2D infinite-horizon
case, whereas lower-dimensional gaps give
normal diffusion. 
Nonetheless,  higher moments of the displacement distribution are
affected by
the small proportion of arbitarily long trajectories in the structure.


To test the analytical arguments, we perform careful numerical simulations of a 
3D periodic Lorentz gas model with spheres of two radii,
which can be varied to obtain different types of diffusive regime. 
In particular, a 
a finite-horizon regime may be obtained by allowing the spheres to overlap;
otherwise, 
gaps of different dimensions can be found. 
Here, results will be given for representative cases in each regime; a detailed
analysis of the model will be given elsewhere.

\begin{figure*}[t]
\includegraphics[scale=0.9]{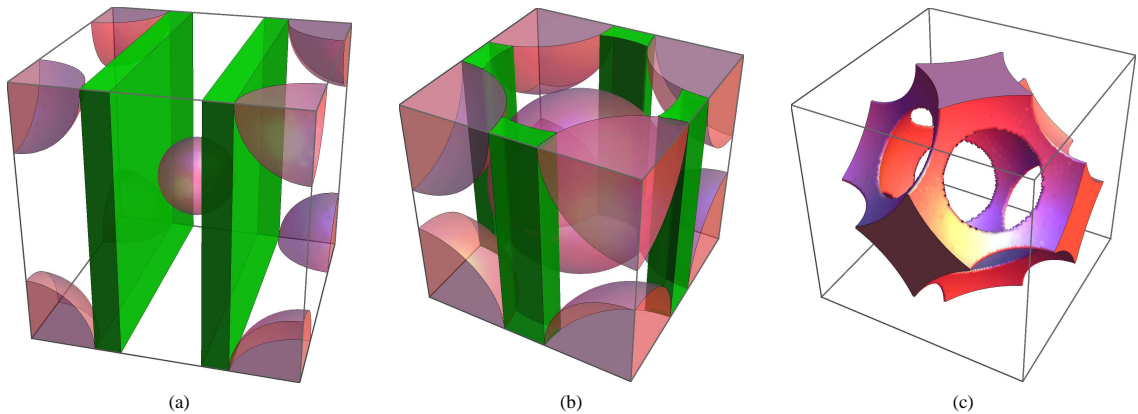}
\caption{(Color online) Spherical scatterers (light color; purple online) and
gaps (dark color; green online) in the 3D periodic Lorentz model discussed in
the text: (a) vertical planar
gaps for $a=0.25$ and $b=0.15$; and (b) vertical cylindrical gaps for $a=0.4$
and $b=0.4$ (a body-centred cubic structure). The gaps are shown in a single
unit cell, but in fact
extend throughout the whole of space. (c) When $a=0.55$ and $b=0.4$,
 the scatterers overlap, leaving an infinite, connected available space for the
particles, which is depicted; for clarity, the spheres are omitted. In this
case, the horizon is finite---there are no gaps in the structure. 
}
\label{fig:gaps}
\end{figure*} 

Finally, we extend the arguments
to higher-dimensional billiards, including 
the class 
of hard-sphere fluids
\cite{SzaszErgodClassicalBilliardBallsPhysicaA1993},
thus providing an approach
to the diffusive behavior of such systems in terms of the geometry
of their
configuration space.

\paragraph{Model and gaps in configuration space:-}

We begin by introducing a simple two-parameter 3D periodic Lorentz
gas model, with which
the different types of diffusive regime can be explored.
The model
consists of a cubic lattice of
spheres of radius $a$, with an additional spherical scatterer of radius $b$ at
the centre of each cubic unit cell, themselves forming  another
(interpenetrating)
cubic lattice. The side length of the cubic unit cell is taken equal to $1$.
By varying the radii $a$ and $b$ of the spheres, a range of models with
different properties can be obtained;  a ``phase diagram'' showing the
possibilities and a detailed study of its properties will be presented
elsewhere.
This is a 3D version of the 2D model
studied in \cite{GarridoGallavottiBilliardCorrelationFnsJSP1994,
SandersFineStructurePRE2005}.  

When $b=0$, we obtain a simple cubic lattice of spherical scatterers.
In this case, we can insert planes parallel to the lattice directions which do
not intersect any obstacles---we call these planar \emph{gaps}.
This remains the case for small enough $b$, 
as shown in
fig.~\ref{fig:gaps}(a).  For $b \ge \frac{1}{2}-a$, however, \emph{all of the planar
gaps are blocked}.  There are still gaps of infinite extent in the structure,
but they are now \emph{cylindrical gaps}, as shown in
fig.~\ref{fig:gaps}(b). 
These are infinitely long tubes which do not intersect any scatterer,
given by the
product of a line with an area; the latter is the projection of the gap 
along the axis of the cylinder.

By tuning $a$ and $b$ appropriately, it is also possible to obtain 
%
an explicit 3D periodic
Lorentz gas
with \emph{finite} horizon.  To do so, we
 allow the scatterers to
overlap, since otherwise constructing such a model is very
difficult.  All adjacent pairs of $a$-spheres overlap when $a > \frac{1}{2}$;
choosing the radius $b$ of the central sphere large enough
then allows us to block \emph{all}
gaps in the structure, giving a finite-horizon model, as shown
elsewhere.  Unlike in the 2D case, in 3D the free space 
between the overlapping scatterers forms an infinite connected network.
Physically,
this can correspond to a
sphere of non-zero radius
colliding with disjoint scatterers.
Note that rigorous results 
on higher-dimensional Lorentz gases assume disjoint scatterers \cite{BalintTothExpDecayMultiDimensional},
and thus do not directly apply to
our model.

\paragraph*{Distribution of free paths:-}

Several approaches to 
the diffusive
properties of
 infinite horizon systems
involve 
the tail of the free-path length distribution,
that is, the proportion $\P(T>t)$ of trajectories, starting from random initial 
conditions in a unit cell, which have a free path length $T$ before colliding 
which is greater
than $t$ \cite{FriedmanMartinDecayVAFLorentzPhysLettA84,
ArmsteadOttAnomDiffnInfHorizPRE2003,
GolseFreePathsLorentzHigherDimensionsM2AN2000}.



%


Consider straight trajectories which emanate in all directions $\v$ from a given
initial condition $\x_0$ lying inside a gap $G$.  Since energy is
conserved at collisions, all particles can be taken to have speed $1$.
The possible
positions $\x_t$ of the trajectories at time $t$ then lie on a sphere $S_t$ of
radius
$t$ and surface
area $S(t)=4 \pi t^2$, centred on $\x_0$.  The proportion $\P(T>t)$ of
trajectories
which have not 
collided during time $t$ is given by the ratio $\P(T>t) := A(t) / S(t)$,
where $A(t)$ is the area of the intersection $I_t := G \cap S_t$ of the gap $G$
with the sphere
$S_t$.


If $G$ is a planar gap, then the intersection $I_t$ is approximately the
 product of a circle of radius $t$ with an interval 
of the same width $w$
as the gap.  Thus 
$A(t) \simeq 2 \pi w t $,
giving the asymptotic behavior $\P(T>t) \sim C/t$ when $t
\to \infty$, where $C$ is a constant. This result was previously found for
a 
simple cubic lattice 
\cite{FriedmanMartinDecayVAFLorentzPhysLettA84,
ChernovStatPropsMultidimPeriodicLorentzGasJSP1994}; a detailed calculation
is given in
\cite{GolseFreePathsLorentzHigherDimensionsM2AN2000}.
%
When $G$ is a cylindrical gap, however, its intersection $I_t$ with the sphere
$S_t$ is
asymptotically the cross-sectional
area $A$ of the cylinder, giving the asymptotics $\P(T>t) \sim C/t^2$.



The tail $\P(T>t)$ of the free-path distribution is strongly related to the
system's diffusive
properties. 
Friedman \& Martin \cite{FriedmanMartinDecayVAFLorentzPhysLettA84} proposed
that the asymptotic decay rate of the velocity autocorrelation function  $C(t)
\defeq \mean{\v(0) \cdot
\v(t)}$ is the same as that of $\P(T>t)$, since $C(t)$ is dominated
by trajectories
which do not collide up to time $t$.
The finite-time diffusion
coefficient $D(t) \defeq \frac{d}{dt} \msd$ is given by $D(t) = \frac{1}
{d} \int_0^t C(s)
ds$, so that $D(t)$ converges, to the diffusion
coefficient $D$, 
only if the velocity
autocorrelation $C(t)$ decays faster than $1/t$ \cite{GaspardBook1998}.



Thus we expect that a 3D periodic Lorentz gas should
exhibit normal diffusion when $\P(T>t)$ decays faster than $1/t$, as is the
case with cylindrical gaps (and when the horizon is finite), but weak
superdiffusion when it decays like $1/t$.  
This is also in agreement with an equivalent condition 
on the moment of the free path distribution between collisions
\cite{BleherStatProps2DInfHorizJSP1992}.


\paragraph*{Numerical results:-}




To test the above hypotheses, we perform careful
numerical simulations of our model to calculate the time-evolution
of the mean-squared displacement $\msd$ in each regime. 
We use a stringent test to distinguish normal from weakly anomalous diffusion:
$\msd / t$ is plotted
as a function of $\log t$ \cite{GeiselThomaeAnomDiffnIntermittentChaoticSystemsPRL1984,
GarridoGallavottiBilliardCorrelationFnsJSP1994,
SandersLarraldeNormalAnomalousDiffusionPolygonalPRE2006}.
Normal diffusion 
corresponds to an
asymptotically flat graph, 
since the logarithmic correction is
 absent, and the diffusion coefficient is then proportional to 
the asymptotic height of the graph.
Weak superdiffusive $t \log t$ behavior for the
mean-squared displacement, on the other hand, gives asymptotic linear
growth 
\cite{SandersLarraldeNormalAnomalousDiffusionPolygonalPRE2006}.  

\begin{figure}
 \includegraphics[scale=0.9]{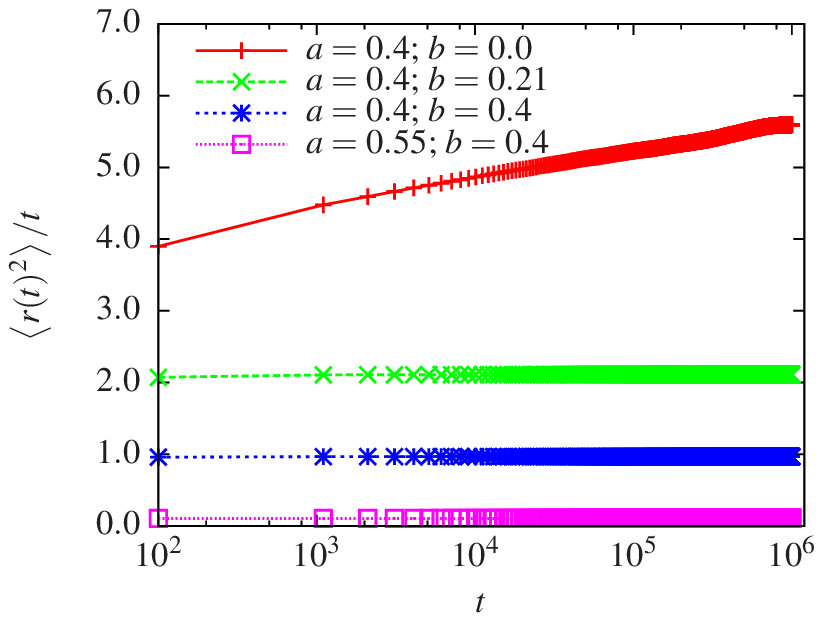}
\caption{(Color online) Linear--log plot of $\msd /
t$ vs.\ $t$ in different
diffusive regimes:
finite horizon ($a=0.55$; $b=0.4$); cylindrical gaps in a 
body-centred cubic lattice ($a=b=0.4$); single large cylindrical gap
($a=0.4$;
$b=0.21$); and simple cubic lattice ($a=0.4$; $b=0.0$) with planar gaps.
Means are taken
over up to $4\times 10^7$ initial conditions; error bars are of the order of
the symbol size. Linear growth (weak superdiffusion) occurs only when there are
planar gaps.}
\label{fig:msd}
\end{figure}

Numerical results are shown in fig.~\ref{fig:msd}.  We see that the
arguments given in the previous section are confirmed:
diffusion is normal, with $\msd \sim t$, when the horizon is finite, and
is weakly superdiffusive, with $\msd \sim t \log t$, when there is a planar
gap. 
Furthermore, the numerics clearly show that \emph{diffusion is
normal} also in the case that there are only cylindrical gaps.  
This is the case
even when the cylindrical gaps
are ``large'', for example when $a=0.4$ and $b=0.21$, when the 
gaps depicted in \ref{fig:gaps}(b) merge to form a single cylindrical gap,
still without any planar gaps in the structure.
%
Thus we conclude that the heuristic arguments correctly predict the type of
diffusion which occurs in these systems.

\paragraph{Gaps in higher-dimensional billiards:-}
Fluids of hard spheres are isomorphic to higher-dimensional chaotic billiard
models, although with
cylindrical instead of
spherical scatterers \cite{SzaszErgodClassicalBilliardBallsPhysicaA1993}.
By extending the above arguments, 
we can hope to obtain information on correlation decay and diffusive properties
for general higher-dimensional chaotic billiards by analysing the gaps in their
configuration space.

To 
define these higher-dimensional gaps,
we 
consider initial positions in a configuration space of dimension $d$,
from which infinitely long non-colliding trajectories emanate along certain
directions. 
We call a connected set of such initial positions for which these non-colliding
trajectories point in the same direction(s) a 
\emph{gap} in configuration space.
Note that it is possible for a given set of initial conditions to have
such trajectories pointing in different, unconnected directions---this is the
case, for example,
%
in fig.~\ref{fig:gaps}(b), where 
there are also 
cylindrical gaps  in a horizontal direction (not shown). In such cases, we 
consider each such set of different directions as a distinct gap.  
For a discussion of higher-dimensional gaps in the context of sphere
packings, see ref.~\cite{ZongDeepHolesFreePlanesReviewBullAMS2002}.

%


As shown above for the 3D case, 
the key geometrical property determining the diffusive behavior of a
system is the dimension of its gaps. We define the
\emph{dimension} of a
gap $G$ to be the 
dimension $g$ of the largest affine subspace which lies
completely within the
gap,
i.e., which does not intersect any scatterer.
In a system with a
$d$-dimensional configuration space, there can be gaps with any
dimension between $1$ and $d-1$, or no gaps at all (finite horizon).

To calculate the tail $\P(T>t)$ of the free-path distribution due 
to such gaps,
we
take coordinates 
$\x := (x_1, \ldots, x_d)$ in the $d$-dimensional configuration space, with
the initial
position at the origin. The
sphere $S_t$ is then given by $\sum_{i=1}^d x_i^2 = t^2$.  Consider a
gap $G$, of dimension $g$.  Inside the gap, there is 
a largest subspace, also of dimension $g$, i.e., it has $g$ 
freely-varying coordinates.
By a rotation of the coordinate system, this subspace can thus be written as
$x_1 = x_2 = \cdots =
x_c = 0$, where $c := d - g$ is the 
\emph{codimension} of the gap, 
giving the number of coordinates in the subspace which are fixed.
The
intersection $I_t = G \cap S_t$ of the
gap with the sphere is thus given by $\sum_{i=c+1}^d v_i^2 = t^2$. This is a
$g$-dimensional sphere, with  surface area $K_g
t^{g-1}$,
where $K_g$ is a dimension-dependent constant. The tail of the free-path
distribution is given by the ratio of the area of intersection $I_t$
to the area of the sphere $S_t$, giving the asymptotics
\begin{equation}
 \P(T>t) \sim Z_c \frac{K_g t^{g-1}}{K_d t^{d-1}} = K t^{-(d-g)} = K t^{-c},
\end{equation}
where $Z_c$ is the $c$-dimensional cross-sectional area of the gap in the
directions orthogonal to the affine subspace, and 
$K$ is an overall
constant.

We thus see that the decay is faster for gaps of smaller dimension (larger
codimension), but it is always eventually dominated by the contribution of
trajectories lying along the gaps.
The dominant contribution to the tail of the free-path distribution, and hence
to the diffusive properties, thus comes from the gap of
largest dimension.

We thus conjecture that $d$-dimensional chaotic, periodic billiard
models 
can generically be expected to
exhibit normal diffusion, at the level of the mean-squared displacement,
provided that the largest-dimensional gap is of dimension less than $d-1$,
that is,
if its codimension is larger than $1$. 


\paragraph{Higher moments:-}
A more sensitive probe of diffusive properties is given by
the growth rates $\gamma(q)$ of the $q$th moments  of the
displacement distribution, $\mean{r^q(t)} \sim t^{\gamma(q)}$,
as a function of the real parameter $q$
\cite{CastiglioneStrongAnomDiffnPhysicaD1999,
ArmsteadOttAnomDiffnInfHorizPRE2003, CourbageZaslavskyTransportBilliardsInfiniteHorizonPRE2008}.  If $\P(T>t)$ decays like 
$t^{-c}$, then long trajectories dominate $\mean{r^q(t)}$  for large $q$,
giving $\gamma(q) = q-c$, while the low moments show diffusive Gaussian
behavior, with $\gamma(q) = q/2$
\cite{ArmsteadOttAnomDiffnInfHorizPRE2003}. A crossover between the two
behaviors
thus occurs at $q=2c$.
If the horizon is finite, then there are no long free flights,
and Gaussian behavior is expected for all $q$.  Higher moments for the finite-horizon Lorentz gas were studied in \cite{ChernovDettmannExistenceBurnettCoefficientsPhysicaA2000}.

\begin{figure}
 \includegraphics{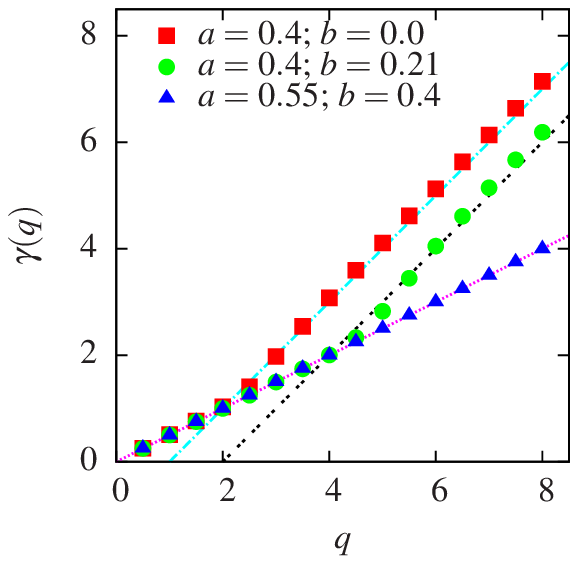}
\caption{Growth rate $\gamma(q)$ of higher moments $\mean{r^q(t)}$ as a
function of $q$; geometries are as in fig.~\ref{fig:msd}. For $a=0.4$ and
$b=0.21$, the means were calculated over $2.4 \times 10^8$ initial conditions,
up to a time $t=10000$, to capture the weak effect of the
cylindrical gaps. The straight lines show the expected Gaussian behavior
($q/2$) and behavior for large $q$ with planar ($q-1$) and cylindrical ($q-2$)
gaps.}
\label{fig:moments}
\end{figure}

The numerical calculation of 
higher moments is difficult, due to the weak effect of free flights
\cite{SandersLarraldeNormalAnomalousDiffusionPolygonalPRE2006}.
Nonetheless, by taking means over a very large number of initial conditions, it
is possible to see the effect 
of the different types of gaps for our 3D Lorentz gas model: as
 shown in 
fig.~\ref{fig:moments}, they are in  agreement with the above argument.
Thus, higher moments can distinguish the subtle effects of different types of
gaps.


In conclusion, we have shown that the diffusive properties of periodic
three-dimensional Lorentz gases, and by extension of higher-dimensional 
periodic billiard models, depend on the highest dimension of 
gap in the configuration space. By introducing a simple
3D model in which each type of diffusive regime occurs, we
showed that if there is a
finite horizon or cylindrical gaps,
then the
diffusion is
normal, whereas planar gaps give weak superdiffusion.  Nonetheless, higher
moments distinguish between different types of gaps.
The concept of infinite horizon is thus no longer 
sufficiently precise for higher-dimensional systems, and must be replaced by
maximal gap dimension. 
In future work we will extend our numerical investigations to higher-dimensional
models.

This work was initiated in the author's Ph.D.~thesis
\cite{SandersPhDThesis2005}. 
He thanks P.~Gaspard 
and R.~MacKay for helpful comments and M.~Henk for useful
correspondence, and 
is grateful to the Erwin Schr\"odinger Institute and the Universit\'e
Libre de Bruxelles for financial support,
which enabled
discussions with N.~Chernov, I.~Melbourne, 
D.~Sz\'asz, I.P.~T\'oth and T.~Varj\'u, and especially T.~Gilbert, who also
read the manuscript critically.
Supercomputing facilities were provided by 
DGSCA-UNAM, and financial support from the DGAPA-UNAM PROFIP programme is also
acknowledged.
The author is grateful to the anonymous referees for interesting comments.


\def\cprime{$'$}

\end{document}